\documentclass[%
 reprint,superscriptaddress,groupedaddress
 amsmath,amssymb,prb,
 aps]{revtex4-2}    
\usepackage{amsmath}
\usepackage{graphicx}
\usepackage{xcolor}
\usepackage{dcolumn}
\usepackage{bm}
\usepackage{bbold}
\usepackage{hyperref}
\usepackage{siunitx}
\usepackage{braket}

\newcommand{\TUdo}{Condensed Matter Theory, Department of Physics, TU Dortmund, 44221 Dortmund, Germany}

\begin{document}

\preprint{arxiv}
\title{Process tensor simulations of non-Markovian open quantum systems with large uncoupled subspaces}
\author{Thomas K. Bracht}
\email{thomas.bracht@tu-dortmund.de}
\affiliation{\TUdo}    
\author{Moritz Cygorek}
\email{moritz.cygorek@tu-dortmund.de}
\affiliation{\TUdo}
\begin{abstract}
\noindent{}Process tensors in matrix product operator representation (PT-MPOs) are a powerful tool to calculate the temporal evolution and multi-time correlations of open quantum systems. While much progress has been made in making the calculation of process tensors for larger system dimensions possible, the numerical cost of the subsequent propagation of the system typically scales as $\mathcal{O}(d^6)$, which renders PT-MPO methods prohibitively expensive for system dimensions above several dozens. Here, we show how PT-MPO methods can be used efficiently when only some of the system states couple to the environment, and how combination with a Taylor expansion method allows for a reduction of the numerical complexity down to $\mathcal{O}(d^3)$. We show that these methods enable efficient simulations of systems with several hundreds of states. This is particularly relevant for solid-state cavity-QED systems with multiple modes, or when calculating frequency-filtered correlation functions by coupling sensor systems to the system of interest. 
\end{abstract}
\maketitle
\section{Introduction}
Quantum systems inevitably interact with their environment, which makes the theoretical description of \textit{open quantum systems} an active field of research. Often, the interaction with the environment leads to effects detrimental to the quantum properties of the system, such as the coherence or entanglement. For many environments, the interaction between system and environment depends not only on the current state of the system, but on past states as well. To accurately predict the dynamics in non-Markovian open quantum systems, sophisticated numerical methods have been developed, such as correlation expansion methods \cite{foerstner2003phonon,vagov2011dynamics}, hierarchical equations of motion \cite{li2012hierachical,tanimura2020numerically,huang2023efficient} or polaron-transformed master equations \cite{wilson2002quantumdot,roy2012polaron,nielsen2021nonmarkovian}.

Several years ago, a numerically complete method for simulating non-Markovian open quantum system was introduced by the process tensor formalism~\cite{JP}. Process tensors~\cite{PT_PRA} in matrix product operator representation (PT-MPOs) represent the complete influence that an environment can have on the system. They are mathematically equivalent to the Feynman-Vernon influence functional~\cite{FeynmanVernon}, but they can be processed much more efficiently due to their matrix product operator (MPO) structure. The following years have seen a series of significant developments in PT-MPO methods directed, on the one hand, towards extending the scope from spin-boson-like environments, e.g., to fermionic~\cite{Abanin} and more general environments~\cite{ACE}, and on the other hand, towards improving their efficiency, e.g., by utilizing time translation invariance to facilitate simulations over millions of time steps~\cite{DnC,LinkStrunz}. 

\begin{figure}
    \centering
    \includegraphics[width=\linewidth]{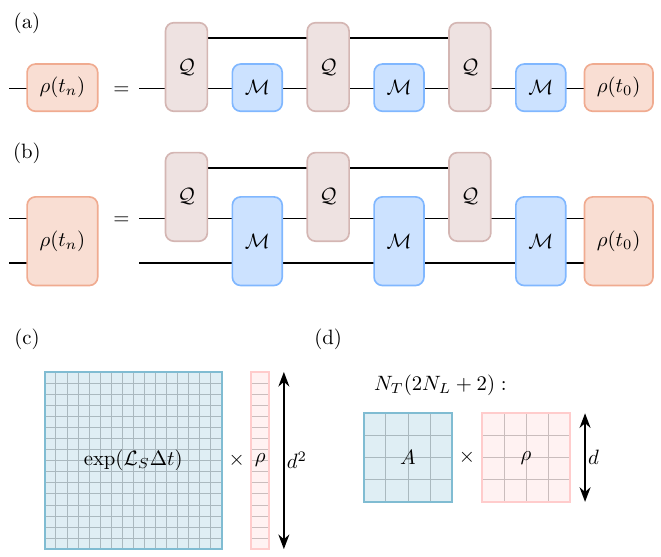}
    \caption{(a) Non-Markovian open quantum system simulation using PT-MPOs. (b) PT-MPO simulation with an additional system subspace not directly coupled to the environment. (c) In Liouville space, the propagation is done by multiplication with a $d^2\times d^2$ matrix. (d) Using the Taylor method, this can be replaced with $N_T(2N_L+2)$ matrix multiplications, where $A=H_S,L_i,L_i^{\dagger},L_i^{\dagger}L_i$ is multiplied to $\rho$ from left or right according to Eq.~\eqref{eq:master_eq}.}
    \label{fig:sketch}
\end{figure}

A current challenge is to tackle the scaling of PT-MPO methods with respect to the dimension $d$ of the Hilbert space of the system of interest. In fact, the computation time of the first PT-MPO algorithms~\cite{JP} scales as $\mathcal{O}(d^8)$. For Gaussian environments, such as the spin-boson model, the contraction scheme can be modified~\cite{Cochin2026} to yield a scaling as $\mathcal{O}(d^4)$.
Moreover, PT-MPOs can be particularly efficiently calculated if degeneracies in the system-environment coupling can be exploited. For example, if the system of interest is composed of two subsystem, one subsystem of dimension $d_c$ which is directly coupled to the environment and one uncoupled subspace \footnote{Here, uncoupled subsystem means that there is no direct coupling term in the Hamiltonian coupling this subsystem to the environment. Yet the uncoupled system is indirectly coupled to the environment via the coupled subsystem. In a numerically complete approach such as ours, this may lead to the build-up of correlations between the uncoupled subsystem and the environment}  of dimension $d_u$, the PT-MPO for the environment can be calculated as if only the coupled subsystem existed [see Fig.~\ref{fig:sketch}b]. The computation time for obtaining the PT-MPO is thus independent of $d_u$.
The analogous concept was utilized in the context of the path integral method QUAPI in Ref.~\cite{cavity_feeding} and its tensor network variant TEMPO in Ref.~\cite{TEMPO}. For PT-MPOs, it is used, e.g., to describe solid-state cavity-QED systems, where a semiconductor quantum dot (QD) with strong non-Markovian coupling to phonons is also strongly coupled to a single-mode microcavity~\cite{bracht2023temperature, Snow2026}.
The same idea also makes it possible to study systems coupled to multiple non-Markovian environments, which describe, e.g., superradiance and collective emission from multiple QDs coupled to their local phonon environments~\cite{Wiercinski1,Wiercinski2}.\\
Yet, one issue remains: for large uncoupled system dimensions $d_u$, obtaining the PT-MPO may not be the limiting factor in the overall PT-MPO method. Propagating the extended density matrix [see Fig.~\ref{fig:sketch}a and b] requires that, in between multiplications with blocks $\mathcal{Q}$ of the PT-MPO, a multiplication with the free system propagator $\mathcal{M}$ is necessary, which is the exponential of the system Liouvillian $\mathcal{M}=e^{\mathcal{L}_S\Delta t}$.
Calculating the matrix exponential using standard methods, such as scaling and shifting of Pad{\'e} approximations~\cite{expm}, requires $\mathcal{O}(d_c^6 d_u^6)$. Even the matrix-vector multiplications needed to propagate the extended density matrix involves $\mathcal{O}(d_c^4 d_u^4)$ operations.
If the free system evolution is purely unitary, the propagator $\mathcal{M}=U \otimes (U^\dagger)^T$ can be decomposed and expressed in terms of the time evolution operator $U=e^{-\frac{i}{\hbar}H_S\Delta t}$. Efficient system propagation would then be possible in $\mathcal{O}(d^3)$. However, many realistic systems require the inclusion of Lindbladian terms to model, e.g., radiative losses.

In this article, we address this unfavorable scaling of PT-MPO simulations with the uncoupled subsystem dimension $d_u$ for general systems, where the free system propagation may be non-unitary. Our approach is based on the Taylor series method of Ref.~\cite{GenericIntegration}, which was originally devised to integrate generic Lindblad master equations describing Markovian open quantum systems. Adapted to PT-MPO simulations, we indeed achieve a scaling of the overall runtime of $\mathcal{O}(d_u^3)$, which makes numerically complete simulations of systems with hundreds of states manageable.

\section{Theory}
\subsection{Process Tensors}
The process tensor framework works as follows: The Liouvillian of the composite system including the environment $\mathcal{L}=\mathcal{L}_S+\mathcal{L}_E$ is split into a part $\mathcal{L}_S$ that only affects the system of interest and the environment part $\mathcal{L}_E$, which also includes the interaction with the system. Formally, the time evolution of total system density matrix $\rho^{tot}$ is given by $\rho^{tot}(t)=e^{\mathcal{L} (t-t_0)} \rho^{tot}(t_0)$. Discretizing the time into steps $\Delta t$, the exponential of the Liouvillian can be decomposed as
\begin{align}
\label{eq:Trotter}
e^{\mathcal{L} t } =&\prod_{l=1}^{n}
e^{\mathcal{L}\Delta t} \approx 
\prod_{l=1}^n \big( e^{\mathcal{L}_S \Delta t} e^{\mathcal{L}_E\Delta t} \big).
\end{align}
Time-dependent Liouvillians can be easily accommodated by choosing the time steps small enough, so that they are essentially constant during $\Delta t$. 
We now introduce labels $\alpha_l$ to enumerate a basis of the reduced system Liouville space, which run over $d^2$ values. Then, the system propagator becomes
a matrix $\mathcal{M}^{\alpha_l,\alpha_{l-1}}=\big(e^{\mathcal{L}_S\Delta t}\big)_{\alpha_l,\alpha_{l-1}}$. Similarly working with an explicit basis (indices $\beta$) of the environment Liouville space for treating the environment propagator is impractical as this space is usually very large or even infinite. However, the tensor network structure of PT-MPOs makes it possible to compress the large number of environment degrees of freedom to the most relevant linear combinations that account for the action of the environment onto the system~\cite{inner_bonds}. This is formally captured by
$\mathcal{Q}^{(\alpha_l,\alpha'_l)}_{d_l,d_{l-1}}=\sum_{\beta_l,\beta_{l-1}}\mathcal{T}_{d_l,\beta_l} \big(e^{\mathcal{L}_E\Delta t}\big)^{\alpha_l,\beta_l,\alpha'_l,\beta_{l-1}}\mathcal{T}^{-1}_{\beta_{l-1},d_{l-1}}$, where $\mathcal{T}$ and $\mathcal{T}^{-1}$ denote rank-reducing compression matrices mapping the complete environment Liouville space to a set of discrete degrees of freedom indexed by $d_l$. 

In practice, the full environment propagator is never constructed explicitly. One rather contracts the PT-MPO from a tensor network, which is composed of small blocks, and which depends on the algorithm used~\cite{TEMPO,ACE}. The typical bond dimension (range of values of indices $d_l$) is denoted by $\chi$ and its magnitude is controlled by convergence parameters, such as a compression threshold $\epsilon$. 
Finally, Eq.~\eqref{eq:Trotter} becomes
\begin{align}
\label{eq:PTsum}
\rho^{\alpha_n}_{d_n}(t_0+ n\Delta t) =&\prod_{l=1}^{n}\bigg(\sum_{\substack{\alpha'_l,\alpha_{l-1},\\d_{l-1}}} 
\mathcal{Q}^{(\alpha_l,\alpha'_l)}_{d_l,d_{l-1}}\mathcal{M}^{\alpha'_l,\alpha_{l-1}} \bigg) \rho^{\alpha_{0}}_{d_{0}} (t_0),
\end{align}
where $\rho^{\alpha_l}_{d_l}$ is an extended density matrix that has a bond index $d_l$ in addition to the system Liouville space index $\alpha_l$.
Eq.~\eqref{eq:PTsum} is summed up by alternately multiplying the extended density matrix with the system propagator $\mathcal{M}$ and the PT-MPO block $\mathcal{Q}$, as visualized in Fig.~\ref{fig:sketch}(a).

Henceforth, we assume a PT-MPO is already calculated, e.g., using one of the methods described in Refs.~\cite{ACE_code,ReviewKeeling,ReviewLacroix}. Our focus is on the scaling of Eq.~\eqref{eq:PTsum} with the system Hilbert space dimension $d$. We note that due to the matrix-vector multiplications with respect to $\alpha$-indices in Eq.~\eqref{eq:PTsum}, the naive scaling of multiplications by $\mathcal{Q}$ and by $\mathcal{M}$ scales as $\mathcal{O}(d^4)$. Yet, for the important case of diagonal or diagonalizable system-bath couplings, which covers the spin-boson model as well as the Rabi model, $\mathcal{Q}$ also becomes diagonal in $\alpha$, so that this multiplication scales as $\mathcal{O}(d^2\chi^2)$.

\subsection{Uncoupled subspaces}
In the context of QUAPI path integral simulations~\cite{cavity_feeding}, it was realized that environment influences can be captured more compactly if several system states couple to the environment identically, i.e., there are degeneracies in the system-environment coupling operator. A special case where this occurs is when the system of interest is composed of two subsystems of dimensions $d_c$ and $d_u$, respectively, where only the first subsystem is directly coupled to the environment while the second is uncoupled and is only indirectly affected by the environment via the coupling through the first subsystem. 

Then, the PT-MPO block $\mathcal{Q}$ is independent of the $d_u$ subspace, which reduces the computation time for multiplication of the extended density matrix to $\mathcal{O}(d^2_u d_c^4\chi^2 )$ for general, off-diagonal subsystem-bath coupling or remains  $\mathcal{O}( d^2\chi^2 )=\mathcal{O}(d_c^2d_u^2\chi^2)$ for diagonal coupling.  Either way, when the uncoupled subsystem becomes large $d_u\gg 1$, the computational bottleneck for summing Eq.~\eqref{eq:PTsum} is the $\mathcal{O}(d_c^4 d_u^4 \chi)$ multiplication with the reduced system propagator $\mathcal{M}$ or the exponentiation of the system Liouvillian, which scales as $\mathcal{O}(d_c^6 d_u^6)$~\cite{expm}. Our strategy to address this bottleneck is to replace the matrix-vector multiplication with $\mathcal{M}$ by an alternative method.

\subsection{Taylor series method}
Recently, it was shown in Ref.~\cite{GenericIntegration} that the integration of a generic Lindblad master equation with $N_L$ Lindblad terms can be efficiently solved in $N_T(2N_L+2)d^3$ operations by expanding the exponential of the Liouvillian in a Taylor series of order $N_T$. This is achieved by replacing the matrix-vector multiplication on the large Liouville space (Fig.~\ref{fig:sketch}c) by a small number of matrix-matrix multiplications on the much smaller system Hilbert space (Fig.~\ref{fig:sketch}d).

Following Ref.~\cite{GenericIntegration}, the scheme works as follows: 
The multiplication of a density matrix $\rho$ with the matrix exponential $e^{\mathcal{L}_S\Delta t}$ is replaced by its $N_T$-order Taylor expansion
\begin{align}
\label{eq:exp_Taylor}
e^{\mathcal{L}_S\Delta t}[\rho] \approx &\sum_{k=0}^{N_T} \frac{\Delta t^k}{k!} \mathcal{L}_S^k[\rho]
=\sum_{k=0}^{N_T} T_k.
\end{align}
The k-th order contribution $T_k$ can be obtained by repeated application of $\mathcal{L}_S[\dots]$ via the iteration
\begin{equation}
\label{eq:iter}
T_0 = \rho, \quad T_k = \frac{\Delta t}{k}\mathcal{L}_S [ T_{k-1} ],
\end{equation}
where the application of $\mathcal{L}_S[\dots]$ with
\begin{align}
\label{eq:master_eq}
\mathcal{L}_S [\rho] =& \frac{i}{\hbar}[\rho, H_S] +
\sum_{i=1}^{N_L}
\gamma_i\bigg(L_i\rho L_i^\dagger -\frac 12\{L_i^\dagger L_i,\rho\}\bigg),
\end{align}
is carried out using $2N_L+2$ matrix-matrix multiplications on Hilbert space, each requiring $d^3$ operations.

Moreover, we note that the Taylor method saves memory in addition to computation time. While the full Liouville space propagator
$\mathcal{M}$ requires the storage of $d^4$ elements, the Taylor method operates on matrices with $d^2$ entries.

\subsection{Taylor series methods for PT-MPO propagation}

To apply the Taylor series method for the $\mathcal{M}$-propagation in Eq.~\eqref{eq:PTsum}, we simply have to apply the same Taylor expansion and iteration as in Eqs.~\eqref{eq:exp_Taylor} and~\eqref{eq:iter}, respectively, but to the extended density matrix $\rho^{\alpha_l}_{d_l}$. This is done for each value of $d_l$ separately. The latter loop can easily be parallelized, but parallelization is turned off for our benchmark calculation for an easier comparison of the different algorithms.

It should be noted that the Taylor series may be terminated by an alternative condition, namely according to the norm of the respective contribution $|T_k|$. Since this contribution is different for different $d_l$, such an approach automatically selects only the most important contributions. For clarity, here we fixed the order of the Taylor expansion in our analysis. 

\section{Results}
\subsection{Dynamics of a solid-state cQED system}
As an example system, we employ a Jaynes-Cummings model where the two-level system part is additionally coupled to a bosonic environment. Such a system can for example be found in the modeling of a semiconductor quantum dot (QD) in an optical cavity \cite{bracht2023temperature}, coupled to longitudinal acoustic (LA) phonons possessing a superohmic spectral density \cite{krummheuer2005puredephasing,lueker2019review}.
A topical example of a non-Markovian open quantum system with a large uncoupled subspace is a solid-state cavity-QED system comprised of a QD embedded in a microcavity. The QD interacts with longitudinal acoustic (LA) phonons via a spin-boson-type interaction \cite{krummheuer2005puredephasing,lueker2019review}. A single mode of the microcavity is described by a truncated Hilbert space of harmonic oscillator eigenstates. Since only the QD directly interacts with the phonons, the coupled system dimension is $d_c=2$ while the size of the decoupled subspace $d_u$ corresponds to the that of the cavity mode.
For large photon numbers, the total Hilbert space dimension rises, while the number of states coupled to the phonon environment stays constant. 

\begin{figure}[t]
	\begin{center}
		\includegraphics[width=\linewidth]{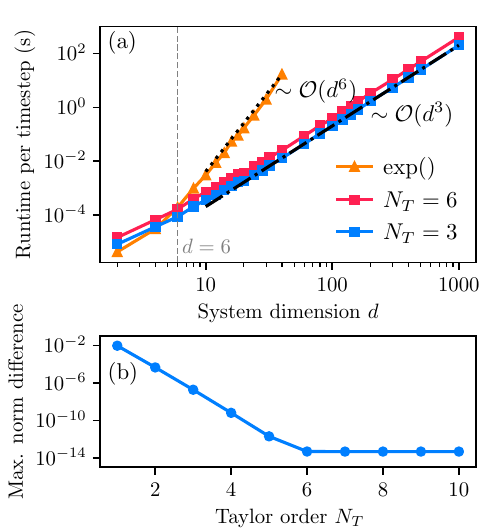}
	\end{center}
	\caption{\label{fig:benchmark} (a) Benchmark of runtime using TLS coupled to single cavity mode with increasing number of cavity photons accounted for, leading to a total dimension $d$.
    (b) Maximum norm difference of density matrix depending on Taylor order $N_T$ compared to usage of exp(), for resonant excitation and subsequent relaxation of two-level system coupled to phonons at $dt=\SI{0.1}{ps}$. Simulations are run on a single core of an Intel Core Ultra 9 285K Processor at factory settings.}
\end{figure}

To simulate the dynamics of this system, we calculate the PT-MPO once for just the QD two-level system coupled to the phononic environment using our own implementation of uniTEMPO \cite{LinkStrunz}. The parameters used for these simulations are listed in the Appendix~\ref{app:a}. 
To propagate the total system including the cavity, we execute the multiplication with $\mathcal{M}$ (see Eq.~\eqref{eq:PTsum}) either by explicitly exponentiating the system Liouvillian or by applying the Taylor series method in Eq.~\eqref{eq:exp_Taylor}. The total runtime is collected for simulations with varying sizes of the truncated bosonic Hilbert space and for two Taylor orders $N_T=3$ and $N_T=6$.
The results are visualized in Fig.~\ref{fig:benchmark}(a), where the runtime per time step is given for system dimensions ranging from $2$ to $1000$. Lines in the double-logarithmic plot indicate the expected scaling behavior. Indeed, from system sizes larger 10, direct exponentiation follows the $\mathcal{O}(d^6)$ trend while the Taylor method retains $\mathcal{O}(d^3)$ scaling. As a result, the numerical cost of propagation becomes feasible for much larger dimensions $d$: While previously a single time step for a system with, e.g., $d=40$ took minutes to compute using direct exponentiation, the Taylor method achieves the same within several milliseconds. Given enough computational resources, propagation of a system up to $d\sim1000$ is possible using this technique.

Notably, the Taylor method outperforms direct exponentiation already for relatively small system sizes of about $d\gtrsim 6$.
This implies that for almost any realistic quantum system apart from simple two- or four-level systems, this method can lead to a considerable speed-up.

While we have confirmed that using the Taylor expansion to describe the exponential leads to considerable gains in numerical efficiency, the question stands whether this reduces the accuracy of the results. PT-MPO methods are numerically exact, meaning the numerical error can be made arbitrarily small depending on convergence parameters and numerical precision. For the Taylor method, the order of the series expansion is the parameter that controls accuracy and convergence.
To analyze this accuracy, we consider a system of size $d=6$ and calculate the norm differences between density matrices obtained using a fixed Taylor order and the respective results obtained using explicit exponentiation. The maximum of the norm differences $\mathrm{max}_t(|\rho_{\text{exp}}(t)-\rho_{N_T}(t)|)$ formed over all time steps is depicted in Fig.~\ref{fig:benchmark}(b) for different Taylor orders $N_T$. 
We find very accurate results starting from $N_T\sim 3$. From $N_T=6$, the numerical error is comparable to the expected numerical error of the Pad{\'e} approximation, close to machine precision.  
\subsection{Frequency-filtered two-time correlations}
\begin{figure}[t]
	\begin{center}
		\includegraphics[width=\linewidth]{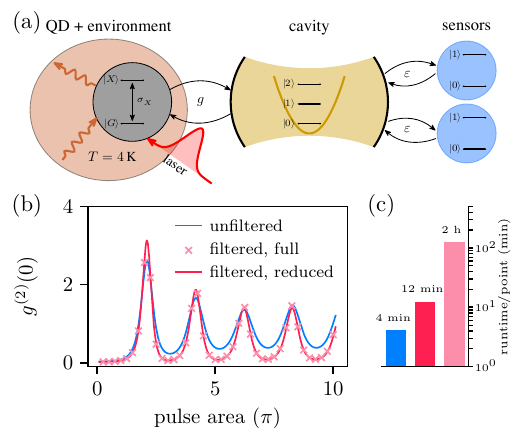}
	\end{center}
	\caption{\label{fig:g2} (a) Two-level system coupled to photonic cavity, which is coupled to two sensor systems allowing to calculate frequency filtered correlations. The total system has a Hilbert space with $d=24$. (b) Calculated $g^{(2)}(0)$ for TLS coupled to LA phonons at $\SI{4}{K}$ under pulsed excitation with varying pulse area, with and without spectral filtering. Red crosses mark data points where all TLS, cavity and sensor states are taken into account, while the red line shows data where only states with a maximum of two total excitations were considered, leading to $d=12$. (c) Histogram showing the runtime of the corresponding simulations for a single data point.}
\end{figure}
One reason why large system Hilbert spaces appear in applications is that certain multi-time correlation functions can be obtained by explicitly modelling the detector as part of the system of interest~\cite{delValle2012,Iles-Smith2026}.
Concretely, spectrally filtered two-photon coincidence signals can be calculated by coupling two additional sensor two-level systems to the cavity mode, where the properties of the filter are set by the energies and Lindbladian decay rates of the sensors [see sketch in Fig.~\ref{fig:g2}(a)].
As in the previous example, we consider a QD modelled as a two-level system, coupled to LA phonons at $T=\SI{4}{K}$. For a detailed list of parameters, see Appendix~\ref{app:b}. Accounting for up to two photons in the cavity, this model leads to a Hilbert space dimension of $d=6$ without sensors and $d=24$ with sensors, while only the two-level system is directly coupled to the environment. Due to the large Hilbert space dimension of the system including the sensors, calculating millions of timesteps necessary to fully resolve dynamical multi-time correlations would be impractically demanding for the conventional system propagation using direct exponentiation of the Liouvillian. However, we now demonstrate that the Taylor method with $N_T=3$ allows one to evaluate the spectrally filtered two-photon coincidences in a reasonable amount of time. 

Here, we excite the QD using Gaussian laser pulses of duration $\sigma=\SI{48}{ps}$ and calculate $g^{(2)}(t,\tau)$ as described in the Appendix~\ref{app:b}. The QD excitations are transferred to the cavity mode as photons, which in turn are partially transferred to the two-level sensor systems. The sensors are tuned to the exciton resonance, with a linewidth of $\SI{19}{\mu eV}$.

The two-time correlation function $g^{(2)}(0)$ is depicted in Fig.~\ref{fig:g2}(b) and shows photon antibunching for pulse areas at even, and pronounced bunching at odd multiples of $\pi$. At even multiples of $\pi$, a photon emission event during the pulse lets the remaining pulse efficiently re-excite the system, leading to the pronounced bunching behavior \cite{fischer2016dynamical}. If filtering is included, low values for $g^{(2)}(0)$ are achieved at even $\pi$ even for elevated pulse areas.

Each $g^{(2)}(t,\tau)$ sample point requires propagating over $\sim 10^7$ time steps. This leads to a total computation time of $\sim 2$~hours for a single data point for the spectrally filtered signal in Fig.~\ref{fig:g2}(c). While this is manageable using the Taylor method, we note that for this particular system even faster simulations are possible by reducing the state space. Accounting only for states where the QD-cavity-sensor systems has at most two excitations reduces the system dimension from $d=24$ to $d=12$, and the computation time from $~\sim 2$ hours to $\sim 12$~minutes. At this reduced dimension, conventional Liouville space propagation would be feasible but tedious, and takes longer ($\sim 7$ hours) than the Taylor propagation on the full system space. We also emphasize that, in order to judge the accuracy of the reduced state space model, the full simulations are necessary, which is only achievable 
with the Taylor method.
\section{Conclusion}
To summarize, we demonstrated a Taylor series method that allows the efficient PT-MPO simulation of systems with large uncoupled subspaces. Concretely, by decomposing the Liouville space propagation into a series of matrix multiplication on the Hilbert space, we achieve a reduction of the computation time from $\mathcal{O}(d^6)$ to $\mathcal{O}(d^3)$. Significant computational savings are found starting from a system Hilbert space dimension of $d\gtrsim 6$, while at the same time the error made by the Taylor expansion is well controlled. Consequently, systems with dimensions of several hundreds are now accessible to a numerically exact treatment, provided the non-Markovian environment couples only to a small subspace. This is particularly relevant for QD-based devices strongly coupled to photonic structures used for non-classical light generation. For these systems, numerically exact predictions of spectrally filtered photon correlation functions~\cite{Liu2025} have so far not been obtainable. 
\section*{Acknowledgments}
We acknowledge the financial support of the Return Program of the State of North Rhine‑Westphalia.
\section*{Code availability}
The implementation of ACE including the Taylor propagation approach is published on github (\url{https://github.com/mcygorek/ACE}). In the parameter files described in Ref.~\cite{ACE_code}, use \texttt{propagate\_Taylor n}, where \texttt{n} is the Taylor expansion order.\\
\appendix
\section{Matrix multiplications for Eq.~\eqref{eq:master_eq}}
Calculating Eq.~\eqref{eq:master_eq} in $2N_L+2$ matrix multiplications is straightforward by defining the non-hermitian effective Hamiltonian
\begin{equation}
    H_{\mathrm{eff}} = H_S - \frac{i}{2}\hbar\sum_{i=1}^{N_L}\gamma_i L_i^{\dagger}L_i.
\end{equation}
$L_i^{\dagger}L_i$ can be pre-computed once, such that calculation of Eq.~\eqref{eq:master_eq} using the given number of matrix multiplication can be achieved by
\begin{equation}
    \mathcal{L}_S[\rho] = -\frac{i}{\hbar}\left(H_{\mathrm{eff}}\rho-\rho H_{\mathrm{eff}}^\dagger\right) + \sum_{i=1}^{N_L}\gamma_i L_i \rho L_i^\dagger.
\end{equation}
\section{Benchmark simulations in Fig.~\ref{fig:benchmark}}\label{app:a}
For the benchmarks shown in Fig.~\ref{fig:benchmark}, we use a TLS coupled to a laser and a photonic cavity, described by
\begin{align}
\begin{split}
    H =& -\frac{\hbar}{2} \left(\Omega^*(t)\ket{g}\bra{x} + \Omega(t)\ket{x}\bra{g}\right)\\
     &+ \hbar g(a\ket{x}\bra{g} + a^{\dagger}\ket{g}\bra{x}) + \hbar\delta_{cx}a^{\dagger}a.
\end{split}
\end{align}
Assuming a Gaussian pulse envelope, $\Omega$ is given as
\begin{equation}
    \Omega(t)= \frac{\Theta}{\sqrt{2\pi}\sigma}e^{-\frac{t^2}{2\sigma^2}},
\end{equation}
where $\Theta$ is the pulse area and $\sigma$ defines the pulse duration.
We further include radiative losses for TLS and cavity using Lindblad operators with $A_i=\{\ket{g}\bra{x}, a\}$ and corresponding rates $\gamma_i=\{\gamma, \kappa\}$, reading
\begin{equation}
    L_{A_i}(\rho) = \gamma_i \left[A_i\rho A_i^{\dagger} -\frac{1}{2}\left\{A_i^{\dagger}A_i,\rho\right\}\right] .
\end{equation}
We apply a $\Theta=\pi$-pulse with $\sigma=\SI{4}{ps}$ resonant to the TLS, the cavity is resonant to the TLS transition, i.e., $\delta_{cx}=0$. The remaining parameters are radiative decay rate $\gamma=1/100\,\si{ps^{-1}}$, cavity coupling $\hbar g=\SI{0.06}{meV}$, cavity loss $\hbar \kappa=\SI{0.12}{meV}$ and a timestep of $dt=\SI{0.1}{ps}$. The system dimension is increased by accounting for an increasing number of cavity states.\\
\section{Calculation of $\mathbf{g^{(2)}(0)}$}\label{app:b}
The calculations in Fig.~\ref{fig:g2} additionally include sensor states for spectral filtering using \cite{delValle2012}
\begin{equation}
    H_{\mathrm{sensor}} = \sum_{i\in\{1,2\}}\hbar\Delta_is_i^\dagger s_i + \hbar\epsilon\sum_{i\in\{1,2\}}s_i^\dagger a + s_i a^\dagger,
\end{equation}
where $\Delta_i$ is the sensor-exciton detuning, and $s_i^{\dagger} (s_i)$ are the sensor transition operators that create (annihilate) an excitation in sensor $i$. 
We use $\sigma=\SI{48}{ps}, \gamma=1/1000\,\si{ps^{-1}}, \hbar g = \SI{0.02}{meV}, \hbar\kappa=\SI{0.052}{meV}$, as well as an exciton-cavity detuning of $\hbar\delta_{cx}=\SI{0.08}{meV}$, sensors resonant to the exciton energy ($\Delta_i$=0) with a linewidth of $\hbar\Gamma=\SI{0.019}{meV}$ (by using $\Gamma$ as a decay rate for transition $s_i$), and cavity-sensor coupling of $\hbar\epsilon=\SI{0.001}{meV}$. Due to the slower dynamics, we use $dt=\SI{1}{ps}$.\\
For calculation of the multi-time correlation function in the main part of the manuscript, we use 
\begin{equation}
    g^{(2)}(t,\tau) = \langle a^{\dagger}_1(t) a^{\dagger}_2(t+\tau) a_2(t+\tau) a_1(t)\rangle,
\end{equation}
where $a_1=a_2=a$ is used for the unfiltered cavity photons, and $a_1=s_1, a_2=s_2$ is used to calculate the results including spectral filtering. Integration along the $t$ axis yields $g^{(2)}(\tau)$, from which $g^{(2)}(0)$ follows as the ratio of events during the zero-delay cycle and the adjacent laser cycle, where the laser repetition time $T=\SI{8}{ns}$ enters \cite{fischer2016dynamical}
\begin{equation}
    g^{(2)}(0) = \frac{\int_{-T/2}^{T/2}d\tau g^{(2)}(\tau)}{\int_{T/2}^{3T/2}d\tau g^{(2)}(\tau)}.
\end{equation}
\bibliography{bibfile}

\end{document}